\documentclass[preprint2]{aastex}

\newcommand{\beq}{\begin{equation}}
\newcommand{\eeq}{\end{equation}}
\newcommand{\bea}{\begin{eqnarray}}
\newcommand{\ena}{\end{eqnarray}}
\begin{document}

 \title{Relative space-time transformations in GRBs}

\author{Remo Ruffini\altaffilmark{1}, Carlo Luciano 
Bianco, Federico 
Fraschetti, She-Sheng Xue}
\affil{ICRA - International Center for Relativistic Astrophysic\\ Physics 
Department, University of Rome ``La Sapienza''}
\affil{Piazzale Aldo Moro 5, I-00185 Rome, Italy.}
\and
\author{Pascal Chardonnet}
\affil{Universit\'e de Savoie - LAPTH LAPP - BP110 - 74941 Annecy-le-Vieux 
Cedex, France}
\altaffiltext{1}{ruffini@icra.it}
%\altaffiltext{5}{chardon@lapp.in2p3.fr}

\begin{abstract}
The GRB~991216 and its relevant data acquired from the BATSE experiment and RXTE and Chandra satellites are used as a prototypical case to test the theory linking the origin of gamma ray bursts (GRBs) to the process of vacuum polarization occurring during the formation phase of a black hole endowed with electromagnetic structure (EMBH). The relative space-time transformation paradigm (RSTT paradigm) is presented. It relates the observed signals of GRBs to their past light cones, defining the events on the worldline of the source essential for the interpretation of the data. Since GRBs present regimes with unprecedently large Lorentz $\gamma$ factor, also sharply varying with time, particular attention is given to the constitutive equations relating the four time variables: the comoving time, the laboratory time, the arrival time at the detector, duly corrected by the cosmological effects. This paradigm is at the very foundation of any possible interpretation of the data of GRBs.
\end{abstract}


In recent years, a large variety of very accurate experimental data, ranging from $\gamma$ rays all the way to the radio band, has been obtained for the afterglows of GRBs, following their first discovery by the BeppoSAX satellite \citep[see e.g.][ and references therein]{c00}.

In the theoretical models of GRBs there are currently three topics under debate:\\
1) The ``internal Shock Model'', introduced by \citet{rm94}, many aspects of which have been developed \citep{px94,sp97,f99,fcrsyn99}. The underlying assumption of this model is that all the variations of GRBs in the range $\Delta t\sim 1$ ms up to the overall duration $T$ of the order of $50$ s are determined by the``inner engine''. The difficulties of explaining the long time scale bursts by a single explosive model has evolved into a class of models assuming an ``inner engine'' with a prolongued activity \citep[see e.g.][ and references therein]{p01}.\\
2) The ``external shock model'', also introduced by \citet{mr93}, is less popular today. It relates the GRBs' light curves and time variations to interactions of a single thin blast wave with clouds in the external medium. There is the distinct possibility, within this model, that ``GRBs' light curves are tomographic images of the density distribution of the medium surrounding the sources of GRBs'' \citep{dm99} \citep[see also][ and references therein]{dcb99,d00}. In this case, the structure of the burst does not come directly from the ``inner engine''.\\
3) In order to decrease the energy requirements of GRBs, the effect of beaming has been advocated \citep[see e.g.][]{my94,dbpt94}. The possibility of inferring its existence from changes in the power-law index of the afterglow is generally considered attractive \citep[see e.g.][]{mr97a,r97,mrw98,pmr98,dc98,sph99,pm99,r99,ha00,gdhl01}.

For the astrophysical nature of the system originating the GRB, a binary system of merging neutron stars has been proposed \citep[see e.g.][]{elps89,npp92,mr92mnras,mr92apj}. Problems occur: a) in the general energetics which cannot exceed $\sim 3\times 10^{52}$ ergs, b) to explain the longer bursts \citep[see][]{swm00,wmm96}, and c) in the observed location of the GRBs' sources in star forming regions \citep[see][]{bkd00}. Alternatively novel classes of astrophysical systems have been postulated, including black hole - white dwarf \citep{fwhd99} and black hole - neutron star binaries \citep{p91,mr97b}, as well as Hypernovae \citep[see][]{p98}, failed supernovae or collapsars \citep[see][]{w93,mw99} and supranovae \citep[see][]{vs98,vs99}.

We take a somewhat intermediate approach by studying the GRB emitted by the process of vacuum polarization around a black hole endowed with electromagnetic structure: the EMBH model. Such a model has the advantage that all the basic intermediate theoretical background, starting from the process of gravitational collapse itself, have been developed. The model can therefore make precise predictions which can be compared with the observations.

We consider a GRB ``prototypical'' case, in order to create a new interpretative paradigm, to be then applied to the observations of other GRBs.  Since some of the best data available, from BATSE \citep{brbr99}, RXTE \citep{cs00} as well as the remarkable accuracy of the Chandra \citep{p00} satellite are available for the GRB~991216, we use it as the prototype. In addition: a) it is one of the strongest observed GRBs. b) it radiates mainly in X- and $\gamma$-rays and less than 3\% is emitted in optical and radio band, and c) a precise value of the slope of the energy emission during the afterglow as a function of time, $n=-1.64$ \citep{tmmgk99} and $n=-1.616\pm 0.067$ \citep{ha00}, has been obtained.

The EMBH model relates the origin of the energy of GRBs to the extractable electromagnetic energy of an EMBH \citep{cr71} via the vacuum polarization process occurring during the gravitational collapse leading to the formation of an EMBH \citep{dr75}. The first step in this theory is the definition of the dyadosphere \citep{rukyoto,prx98}, an extended region outside the EMBH horizon formed of an optically thick plasma of electron-positron pairs and radiation whose energy $E_{dya}$ is related to the mass $ \mu = M/M_\sun$ and electromagnetic parameter $ \xi = Q/\left(M\sqrt{G}\right)$ of the EMBH by the relation:\\
\beq
\displaystyle{
E_{dya} = \frac{Q^2}{2 \; r_{+}} \left(1 \; - \; \frac{r_{+}}{r_{ds}} 
\right)
\left[ 1 \; - \; \left(\frac{r_{+}}{r_{ds}}\right)^2 \right] ,
}
\eeq 
where  $r_{+} =  1.47 \times 10^5 \mu ( 1 \;+ \; \sqrt{1 - \xi^2} ) $ is the horizon radius and  $ r_{ds} = 1.12 \times 10^8 \sqrt{\mu \xi} $ is the dyadosphere radius and, as usual, $M$ and $Q$ are the mass-energy and charge of the EMBH and $G$ is the Newton constant of gravity. 

The evolution of this pair-electromagnetic plasma leads to the formation of a sharp pulse (the PEM  pulse) that very rapidly reaches a Lorentz gamma  factor of $10^2$ and higher.  The subsequent interaction of this pulse with the baryonic matter of the remnant, left over from the gravitational collapse of the protostar, and with the interstellar medium (ISM) leads to the different eras of the GRBs. It is useful to parametrize the baryonic mass $M_B$ of the remnant by introducing the dimensionless parameter $B$:
\beq
M_B c^2 = B E_{dya} .
\eeq

The confrontation of the theoretical model with the observational data allows an estimate for the values of the EMBH parameters. It also allows us to probe the density of the baryonic material in the remnant, in the ISM as well as in the stellar distribution within a few parsecs of the EMBH \citep[see][]{lett2,lett3}.

The first step in this process is the establishment of the first set of constitutive equations relating:\\
a) The comoving time of the pulse ($\tau$): the evolution of the thermodynamical quantities (density, temperature) are computed using this time.\\
b) The laboratory time ($t$) defined by an inertial reference frame in which the EMBH is at rest.\\
c) The arrival time ($t_a$): the laboratory time at which light signals from the source reach a distant observer at rest in the laboratory frame. The zero of the arrival time has been chosen to coincide with the arrival of the light signals from the moment of formation of the EMBH\\
d) The arrival time at the detector ($t_a^d$): this is the arrival time taking into account the cosmological redshift of the GRB source. We have
\begin{equation}
t_a^d = t_a \left(1+z\right),
\label{taddef}
\end{equation}
where $z$ is the cosmological redshift of the GRB source \citep{lett6aa,bcfjrx01}. In the case of GRB 991216 we have $z\simeq 1.00$.

The mutual relations of these four times with the radial coordinate in the laboratory frame is the subject of this letter. We first give emphasis to a basic feature of the arrival time determination. For signals emitted by a pulse moving with velocity $v$ in the laboratory frame, we have:

\begin{equation}
\Delta t_a  = \left( {t_0  + \Delta t + \frac{{R_0  - r}}{c}} \right) - \left( {t_0  + \frac{{R_0 }}{c}} \right) = \Delta t - \frac{r}{c} ,
\label{taintr}
\end{equation}
where $\Delta t_a$ ($\Delta t$) is the time interval in arrival (laboratory) time, $R_0$ is the distance of the observer from the EMBH, $t_0$ is the laboratory time corresponding to the gravitational collapse, and $r$ is the radius of the expanding pulse at the time $t=t_0 + \Delta t$.

For simplicity we abbreviate the interval notation $\Delta t_a$ ($\Delta t$) by $t_a$ ($t$). Eq.(\ref{taintr}) can then be rewritten as:

\begin{equation}
t_a  = t - \frac{r}{c} = t - \frac{{\int_0^t {v\left( {t'} \right)dt'}  + r_{ds} }}{c} ,
\label{tadef}
\end{equation}
where the dyadosphere radius $r_{ds}$ is the value of $r\left(t=0\right)$. We consider only the photons emitted along the line of sight, since the spreading due to the angular dependence and to the thickness of the pulse is negligible \citep[see][ for details]{brx01}. The solution of Eq.(\ref{tadef}) neglecting $r_{\rm ds}$:

\begin{equation}
t_a  = t - \frac{{v_0 }}{c}t - \frac{1}{2}\frac{a}{c}t^2  -  \ldots , 
\label{taex}
\end{equation}
is in general highly nonlinear.

If and only if $v$ is constant and $v\simeq c$, Eq.(\ref{tadef}) can be rewritten, neglecting $r_{\rm ds}$, as:

\begin{equation}
t_a  \simeq t\left( {1 - \frac{v}{c}} \right) = t\frac{{\left( {1 - \frac{v}{c}} \right)\left( {1 + \frac{v}{c}} \right)}}{{\left( {1 + \frac{v}{c}} \right)}} \simeq \frac{t}{{2\gamma ^2 }} .
\label{taapp}
\end{equation}
It is clear that the knowledge of $t_a$, which is indeed essential for any physical interpretation of GRB data, depends on a definite integral whose integration limits extend from the gravitational collapse to the time $t$ relevant for the observations, see Eq.(\ref{tadef}). Such an integral is {\em not} generally expressible as a simple linear relation or even by any explicit analytic relation since we are dealing with processes with variable Lorentz gamma factors of unprecedented magnitude and time variability. Most studies have adopted an approximation of the kind given in Eq.(\ref{taapp}) \citep[see e.g.][]{fmn96}. We instead use Eq.(\ref{tadef}). The adoption of Eq.(\ref{taapp}) misses a crucial feature of the GRB process and it leads to a subversion of the space-time relations in GRBs, with a wide range of consequences: all theoretical computations on the power-law indexes of the afterglow are affected. Specific illustrative examples, pointing out these differences, are shown in the next paragraphs \citep[see][ for details]{lett6aa}. 

The book-keeping of the four different times and corresponding space variables must be done carefully in order to keep the correct causal relation in the time sequence of the events 
involved. This will have also important consequences in the supernovae-GRBs correlation \citep[see][]{lett3}

The second set of constitutive equations are the full non-linear relativistic hydrodynamic equations of energy and momentum conservation, to be solved together with the rate equation for the $e^{\pm}$ plasma. The computations carried out semi-analytically in Rome have been validated by the full numerical computations performed using Wilson's codes at Livermore \citep[see][]{rswx99,rswx00,lett6aa}.

We have integrated both sets of constitutive field equations given in \citet{rswx99,rswx00,brx00}, for the source GRB~991216. Correspondingly, we have obtained the parameter values presented in \citet{lett2}: $E_{dya} \simeq 10^{53}$ ergs  and   $B = 4 \times 10^{-3}$. These values correspond to any of the following pairs of values for the EMBH mass and charge to mass ratio ($\mu,\xi$) =(22.3, 0.1); (10.0,0.15); (5.5,0.2).

\begin{figure}
\plotone{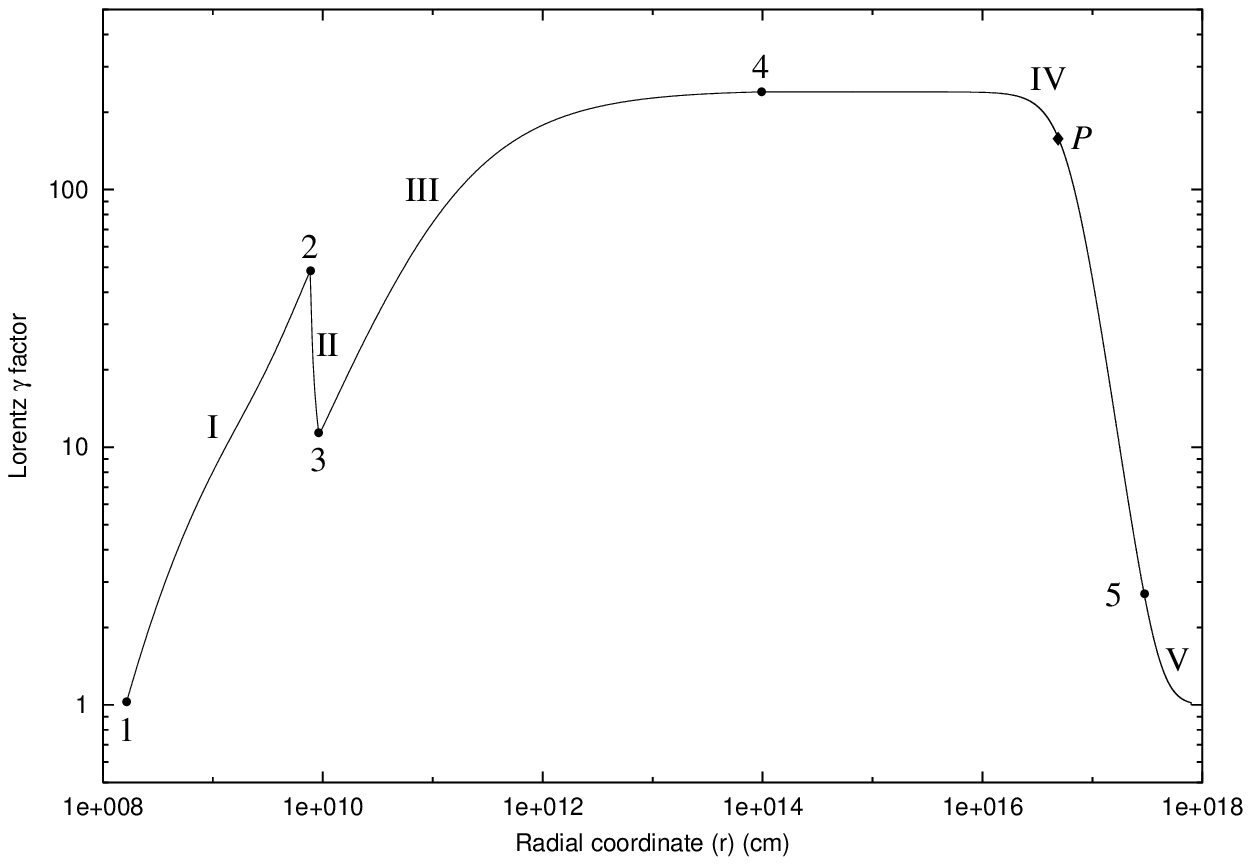}
\caption{The theoretically computed Lorentz gamma factor for the parameter values $E_{dya}=9.57\times 10^{52}$ erg, $B=4\times 10^{-3}$ is given as a function of the radial coordinate in the laboratory frame. The corresponding values in the comoving time, laboratory time and arrival time are given in Tab.~\ref{tab1}.  The different eras, indicated by roman numerals, are illustrated in the text, while the points 1,2,3,4,5 mark the beginning and end of each of these eras. The point P marks the maximum of the afterglow flux \citep[see][]{lett2}. At point 4 the transparency condition is reached.}
\label{gammaz}
\end{figure}

\begin{figure}
\plotone{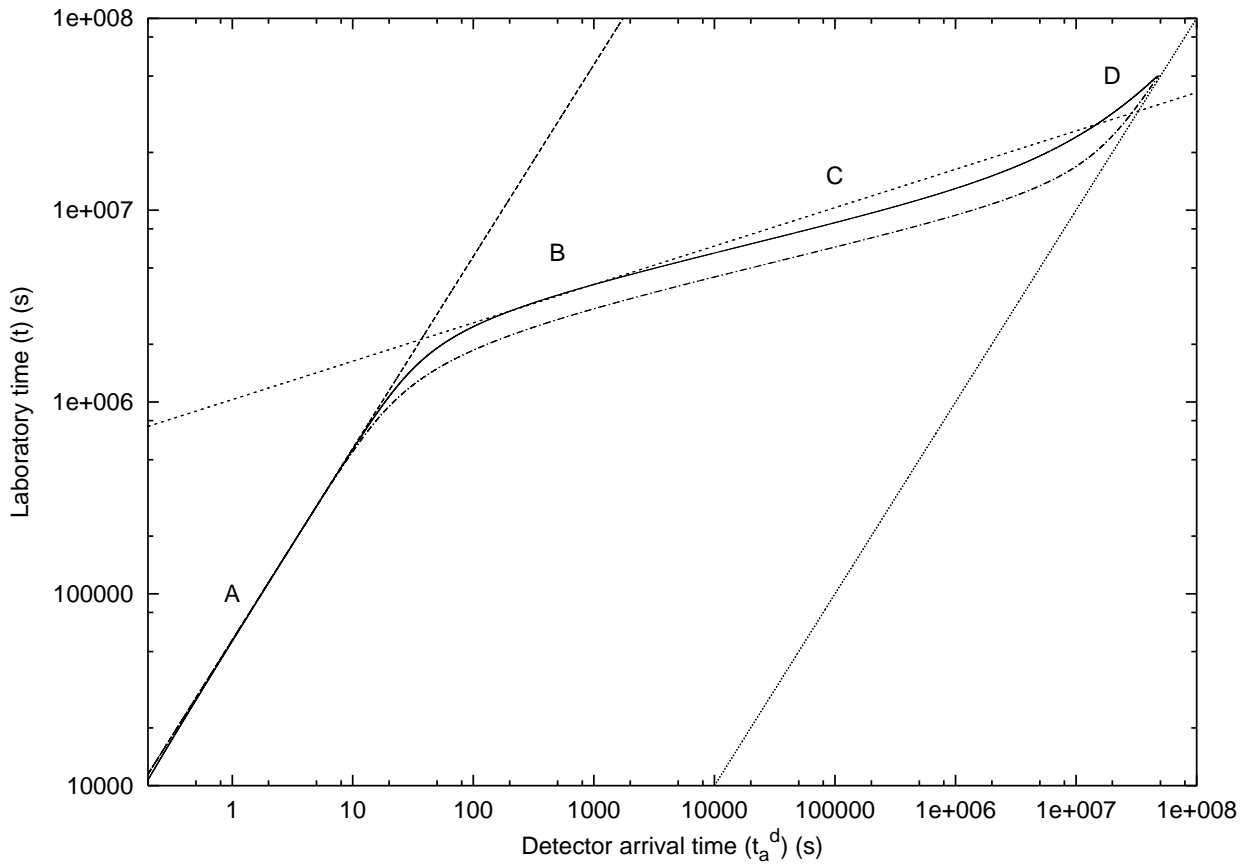}
\caption{Relation between the arrival time ($t_a^d$) measured at the detector, and the laboratory time ($t$) measured by an observer at rest in an inertial frame also at rest with the GRB source. The solid curve is computed using the exact formula given in Eq.(\ref{tadef}), incorporating as well the cosmological effects given in Eq.(\ref{taddef}), while the dashed-dotted curve is the corresponding line obtained using the approximate formula given in Eq.(\ref{taapp}), computed with the time varying $\gamma$ given in Fig.~\ref{gammaz}. This is the approximation used by {\em all} recent works on GRBs. The difference is very conspicous, if one takes into account that the diagram is in logaritmic scale, and has basic consequences on the astrophysical scenario \citep[see e.g.][]{lett2,lett3}. We distinguish four different phases. 
{\bf Phase A}: There is a linear relation between $t$ and $t_a^d$ (dashed line). 
{\bf Phase B}: There is an ``effective'' power-law relation between $t$ and $t_a^d$ (dotted line). 
{\bf Phase C}: No single analytic formula holds, the relation between $t$ and $t_a^d$ can only be analysed patchwise and has to be directly computed by the integration of the complete equations of energy and momentum conservation.
{\bf Phase D}: As the Lorentz factor approaches $\gamma=1$, the relation between $t$ and $t_a^d$ asymptotically goes to $t=t_a^d$ (light gray line). Details given in \citet{lett6aa}.}
\label{tvstazC}
\end{figure}

Crucial to any GRB data interpretation is the relation of the Lorentz gamma factor to the radial coordinate of the source in the laboratory frame and the corresponding values of the above four time parameters. In Fig.~\ref{gammaz} the gamma factors for the different eras are given as a function of the radial coordinate of the source in the laboratory frame. Correspondingly we present in Fig.~\ref{tvstazC} the relation between the laboratory time and the detector arrival time for the source GRB~991216. The highly nonlinear behaviour is obvious, and the different results obtained from the use of Eqs.(\ref{taex},\ref{taapp}) are clearly visible. Details are given in \citet{lett6aa}.

In Tab.~\ref{tab1}, for each successive ``era" and for one very significant event, we give the initial-final values of the gamma Lorentz factor, the four time parameters mentioned above, as well as the radial coordinates in the laboratory frame.

We then have:

1) Era I. The pair-electromagnetic plasma, initially at $\gamma=1$, expands away from the EMBH horizon and from the dyadosphere as a pulse, the PEM pulse.  In the comoving frame the thickness of the pulse increases during the expansion, but the Lorentz contraction in the laboratory frame exactly balances this expansion so that, in the laboratory frame, a constant thickness approximation can be adopted for the burst \citep{rswx99}.  The expansion of the PEM pulse occurs in a region of very low baryonic contamination with density $\rho_B\ll 10^{-9} g/cm^{3}$ \citep{CNR}.  The final Lorentz factor and space time parameters are given for point 2 in Tab.~\ref{tab1}.

2) Era II.  While the PEM pulse is still optically thick it reaches the remnants left over by the gravitational collapse of the progenitor star. The engulfment of this baryonic material induces by conservation of energy and momentum a drastic reduction  in the $\gamma$-factor \citep{rswx00}. The amount of baryonic matter in the remnant has been fixed by the determination of the parameter $B$ in the fitting of the afterglow data \citep[see][]{lett2}. Since these data contain important direct information on the progenitor star, we report in Tab.~\ref{tab2} some specific values of the parameters corresponding to selected values of the EMBH masses: they include the radius and thickness of the remnant as well as the density of baryonic matter. The results are largely independent of the thickness $\Delta$ of the remnants \citep{rswx00}; they depend crucially only on the value $B$. The final Lorentz factor and space-time parameters are given for point 3 in Tab.~\ref{tab1}.

3) Era III.  A new pulse is formed composed of electron-positron pairs and baryons and electrons of the remnant material, a PEMB pulse. Since the opacity from Thomson scattering consequently increases,  the process of self-acceleration of the burst continues to even larger values of the Lorentz gamma factor, which may reach values up to $10^3$--$10^4$ in some sources \citep{rswx00}. In the present case of GRB~991216, the maximum value reached is $239.6$. It is remarkable that the constant thickness approximation for the pulse is still valid \citep{rswx99,rswx00}. This era ends at point 4 as the condition of transparency is reached.  At that point, what we define here as the proper gamma-ray-burst (P-GRB) is emitted \citep{rswx99,rswx00,brx00}. This definition is assumed in order to distinguish the overall GRB phenomena from the specific emission occurring as the moment of transparency is reached. The shape, the intensity and the duration of this P-GRB, also called in the literature the elementary spike \citep{brx00}, strongly depend on the baryonic matter content \citep{brx01,lett6aa}. The final Lorentz factor and space-time parameters are given for point 4 in Tab.~\ref{tab1}.

These first three eras have no counterparts in earlier models, since no detailed description of the early phases of GRB has been attempted.

4) Era IV.  The accelerated baryonic matter expands still as a constant thickness pulse (ABM pulse) at ultra-relativistic velocities and engulfs baryons and electrons from the interstellar medium (ISM), which is assumed to have a constant number density $n_{ism}$ of 1 proton$/cm^{3}$. We have assumed that all internal energy made available by the relativistic conservation of energy and momentum is radiated away in the afterglow, mainly in $\gamma$ ($\sim 90\%$) and X-rays ($\sim 10\%$), and a few percent in the optical and radio emission \citep[see][]{ha00}. We have used the ``fully radiative case'' condition \citep[see e.g.][]{p99}, considering only the leading contribution of the head-on flux. We have neglected the spreading due to off-axis emission, considered e.g. in \citet{fcrsyn99} and have given physical reasons for neglecting such contributions \citep{lett6aa}. At any specific time, the total flux, and consequently the bolometric luminosity of the afterglow, is fixed by the above requirements. The detailed spectral distribution depends on the dominant radiative process of the internal energy. A variety of such process have been considered in the literature and their results within the context of our model are given in \citet{lett6aa}. In Fig.~\ref{gammaz} we show how, during this era, the Lorentz gamma factor first coasts to a constant value and then rapidly decreases, going from $ \gamma = 239.6$ to $\gamma = 2.7$.  Most important is that the point P where  $ \gamma \simeq 160.2 $ corresponds to the peak of the afterglow \citep[see][]{lett2}. Beyond this point P the slope of the afterglow flux, as a function of arrival time, approaches the power law index $n=-1.6$ in perfect agreement with the observations of RXTE and Chandra \citep[see][]{lett2}. The final Lorentz factor and space-time parameters are given for point 5 in Tab.~\ref{tab1}. It is important to emphasize that this power law index results from the combination of three critical assumptions: a) the emission occurring in a ``fully radiative'' regime, b) the condition of spherical simmetry, and c) the constancy of the ISM density. Earlier results relevant to this treatment can be found in \citet{s97} and in \citet{dcb99} \citep[see, for comparison and contrast,][]{lett6aa}.

5) Era V.  This is the transition to the relativistic and nonrelativistic regimes. This era is more complex. It contains two successive sub-eras, one with a power-law index of the energy emitted in the afterglow as a function of the detector arrival time $n=-1.36$, corresponding to a still relativistic era ($1.1\le\gamma\le 2.7$), and a final one approaching the pure newtonian regime, with $n=-1.45$ and $\gamma < 1.1$. Comparison and contrast with existing slopes in the literature \citep[see e.g.][]{v97} are presented in \citet{lett6aa}. No data of GRB~991216 are available for checking the thoretical predictions of this last era.

In conclusion we see from Tab.~\ref{tab1} and Fig.~\ref{gammaz} the remarkable and perfectly reasonable results that a motion of the pulse corresponding to a displacement of $9.692\times 10^{13}$ cm  will correspond to an arrival time interval of $1.360\times 10^{-1}$ sec, leading to what has been called apparent superluminal behavior. Similarly, on a larger scale, a displacement of the pulse by $2.958\times 10^{17}$ cm will correspond to an increment of $2.439\times 10^5$ sec in arrival time, leading again to apparently superluminal behavior.

From the above results we are ready to express the relative space-time transformation (RSTT) paradigm: the necessary condition in order to interpret the GRB data, given in terms of the arrival time at the detector, is the knowledge of the {\em entire} worldline of the source from the gravitational collapse. In order to meet this condition, given a proper theoretical description and the correct constitutive equations, it is sufficient to know the energy of the dyadosphere and the mass of the remnant of the progenitor star \citep[see][]{lett2}. The application of this RSTT paradigm will have important consequences for the interpretation of the burst structure (IBS), leading to a new paradigm \citep[see][]{lett2}, as well as for the GRB-supernova correlation \citep{lett3}.

\acknowledgments

We thank three anonymous referees for their remarks, which have improved the presentation of this letter.

\clearpage

\onecolumn

\clearpage

\begin{deluxetable}{clllllr}
\tablecaption{Lorentz factors for selected events and their space-time 
coordinates. \label{tab1}}
\tablewidth{0pt}
\tablehead{
\colhead{Point} & 
\colhead{$r$ (cm)} & 
\colhead{$\tau$ (s)} &
\colhead{$t$ (s)} &
\colhead{$t_a$ (s)} &
\colhead{$t_a^d$ (s)} &
\colhead{$\gamma$}
}
\startdata
1 & $1.610\times10^8$ & $0.0$ & $0.0$ & $0.0$ & $0.0$ & $1.00$\\
2 & $7.659\times10^9$ & $1.985\times 10^{-2}$ & $2.580\times 10^{-1}$ & $1.846\times 10^{-3}$ & $3.692\times 10^{-3}$ & $48.38$\\
3 & $9.153\times10^9$ & $2.292\times 10^{-2}$ & $3.089\times 10^{-1}$ & $2.780\times 10^{-3}$ & $5.559\times 10^{-3}$ & $11.38$\\
4 & $9.692\times 10^{13}$ & $14.23$ & $3.295\times 10^3$ & $6.805\times 10^{-2}$ & $1.361\times 10^{-1}$ & $239.6$\\
P & $4.863\times 10^{16}$ & $7.784\times 10^3$ & $1.653\times 10^6$ & $11.86$ & $23.72$ & $160.2$\\
5 & $2.958\times 10^{17}$ & $1.082\times 10^6$ & $9.989\times 10^6$ & $1.2195\times 10^5$ & $2.439\times 10^5$ & $2.7$\\
\enddata

\end{deluxetable}

\begin{deluxetable}{crrrrrrrrrrr}
% \tabletypesize{\scriptsize}
\tablecaption{Baryonic matter of the remnant.}
\tablewidth{0pt}
\tablehead{
\colhead{M \tablenotemark{a} } & 
\colhead{ $\xi $ } & 
\colhead{$r_{ds}$\tablenotemark{b} } & 
\colhead{$r_{shell}$\tablenotemark{b} }   &
 \colhead{$\Delta_{shell}$\tablenotemark{b} } &
 \colhead{$\rho$\tablenotemark{c}}}   

  % \colhead{$E_a$\tablenotemark{d}}}

\startdata
  $22.3$ &  $0.10$  &$ 1.67 \times 10^8 $   & $8.36 \times 10^9$  &  $ 1.67 
\times 10^9 $     &  $0.30$   \\
  $10.0$ & $0.15$ & $ 1.37 \times 10^8 $   & $6.86 \times 10^9$  &  $ 1.37 
\times 10^9 $     &  $0.55$   \\
  $5.5$  &  $0.20$   &$ 1.17 \times 10^8 $   & $5.8 \times 10^9$  &  $ 1.17 
\times 10^9 $     &  $0.90$   \\
\enddata

\tablenotetext{a}{in solar mass }
\tablenotetext{b}{in the laboratory frame in cm }
\tablenotetext{c}{in $ g.cm^{-3}$ }
 \label{tab2}

\end{deluxetable}

\end{document}